\documentclass[preprintnumbers,amsmath,amssymb,showpacs, twocolumn]{revtex4}

\usepackage{graphicx}
\usepackage{dcolumn}
\usepackage{bm}

\begin{document}

\title{  Capacity of a  simultaneous  quantum secure direct
communication scheme between the central party and other $M$
parties}

\author{Ting Gao $^{1,2}$,  Feng-Li Yan  $^{2,3}$,  Zhi-Xi Wang $^4$, You-Cheng Li $^{2,3}$}

\affiliation { $^1$ College of Mathematics and Information
Science, Hebei Normal
University, Shijiazhuang 050016, China\\
$^2$ CCAST (World Laboratory), P.O. Box 8730, Beijing 100080, China\\
$^3$ College of Physics  and Information Engineering, Hebei Normal University, Shijiazhuang 050016, China \\
$^4$  Department of Mathematics, Capital Normal University, Beijing 100037, China\\
}

\date{\today}
\begin{abstract}

 We analyze the  capacity of  a  simultaneous quantum
secure direct communication  scheme between the central  party and
other $M$  parties via $M+1$-particle GHZ states and swapping
quantum entanglement.  It is shown that  the encoding scheme
should be secret if other $M$ parties  wants to transmit $M+1$ bit
classical messages to the center party secretly. However when the
encoding scheme is announced publicly, we prove that the capacity
of the scheme in transmitting the secret messages is 2 bits, no matter how big $M$ is.\\

\end{abstract}


\maketitle

Quantum cryptography or quantum key distribution (QKD) utilizing the
features  of quantum mechanics, such as uncertainty principle,
quantum correlations and non-locality, is one of the most promising
applications of quantum information science and is considered to be
the safest system to enable the provable secure distribution of
private information among the authorized parties. It has attracted
widespread attention since the seminal work on QKD by Bennett and
Brassard (BB84) \cite {BB84}.  As shown in this protocol, two remote
authorized users, Alice and Bob, can establish a shared secret key
with nonorthogonal polarized single photons. The quantum non-cloning
theorem \cite {WZ} and the tradeoff between information gain and
disturbance \cite {NC} guarantee  strongly the unconditionally
secure of the BB84 protocol \cite {ShorPreskill}. In virtue of
correlation of an Einstein-Podolsky-Rosen (EPR) pair, the maximally
entangled two-particle state, Ekert proposed a QKD protocol, in
which the Bell inequality is used for detecting the eavesdropper Eve
\cite {Ekert91}. Since then
 there have been various  theoretical proposals \cite {B92,BBM92,
 LongLiu,XueLiGuo,DengLong2003,DengLong2004}, which have been summed up in a review paper \cite {GRTZ}.

Recently Shimizu and Imoto \cite {SIpra60, SIpra62} and Beige et
al \cite {Beige} came up with a novel quantum secure direct
communication (QSDC) schemes, which allows information to be
transmitted in a deterministic secure way. Bostr$\rm \ddot{o}$m
and Felbinger presented a QSDC protocol called the "ping-pong
protocol"  based on entanglement and two-way communication \cite
{BF}. W$\rm \acute{o}$jcik investigated the security of the
ping-pong protocol in a noisy quantum channel \cite {Wojcik}. The
ping-pong protocol was modified by Cai and Li  with single photons
\cite {CaiLi}. Deng et al introduced  two QSDC schemes, one using
EPR pair block \cite {DLL} and the other based on polarized single
photons \cite {DLpra69}.  A QSDC scheme with quantum superdense
coding \cite {WDLLLpra71} was designed by Wang et al. Some ideas
in entanglement swapping has been exploited for QSDC by Man, Zhang
and Li \cite {ManZhangLi}. Zhu, Xia, Fan and Zhang put forward  a
 QSDC protocol based on secret transmitting order of
particles \cite {ZhuXiaFanZhang}. Cao and Song proposed a QSDC
protocol with W state \cite {CaoSong}. In virtue of quantum
teleportation and entanglement swapping we suggested several QSDC
schemes \cite {YZ,Gaozna,GYWcp,GYWijmpc,GYWnc,GYWjpa2,GYWcpl}.
Recently, the QSDC has been extended to multiparty quantum secret
report by Deng et al \cite {DengLiLiZhouLiangZhou}.

In Ref. \cite {GYWcpl} we  presented a simultaneous quantum secure
direct communication scheme between the central party and other $M$
parties, which utilizes shared $M+1$-particle
Greenberger-Horne-Zeilinger (GHZ) states and entanglement swapping
between communicating parties. For the sake of simplicity in \cite
{GYWcpl}, we mainly paid  our attention to the protocol in the case
of $M=3$, which is briefly described as follows. Suppose that a lot
of four-party  GHZ states (GHZ quadruplets)
\begin{equation}\label{2GHZ4}
   |GHZ\rangle=\frac{1}{\sqrt{2}}(|0000\rangle+|1111\rangle)
\end{equation}
are shared by the  three spatially separated senders, Alice, Bob
and Charlie and a remote center party (receiver) Diana. After
ensuring the safety of the quantum channel, Alice, Bob and Charlie
encode
 secret classical bits by applying predetermined unitary operations (encoding scheme) on GHZ quartets. Then
 Alice, Bob, Charlie and  Diana make  Bell state
 measurements on each GHZ quartet pair composed  of one GHZ state
 encoded by the senders
and another original GHZ quartet. Based on  Alice, Bob and
Charlie's Bell state measurement results announced publicly and
Diana's outcome which is kept in secret, Diana can infer the
secret messages of the senders.

In \cite {GYWcpl} we did not state clearly whether the encoding
scheme is secret or not. However, it is the encoding scheme that
determines the capacity of the communication scheme. In present
paper we will investigate this problem and show that  the encoding
scheme should be secret if other $M$ parties  wants to transmit
$M+1$ bit classical secret messages to the center party; but when
the encoding scheme is announced publicly,  the capacity of the
scheme in transmitting the secret messages is 2 bits, no matter how
big $M$ is.

Let us  analyze the capacity of the scheme in transmitting the
secret message when the encoding scheme is announced publicly.
Without loss of generality, we consider the case of $M=3$.

The encoding scheme used in the protocol \cite {GYWcpl} is that
Alice  applies the unitary operations
\begin{equation}\label{2operation1}
\begin{array}{cc}
 \sigma^A_{00}=I=|0\rangle\langle 0|+|1\rangle\langle1|, & \sigma^A_{01}=\sigma_x=|0\rangle\langle1|+|1\rangle\langle0|, \\
 \sigma^A_{10}={\rm i}\sigma_y=|0\rangle\langle1|-|1\rangle\langle0|, &
\sigma^A_{11}=\sigma_z=|0\rangle\langle0|-|1\rangle\langle1|
\end{array}
\end{equation}
 on GHZ quadruplets,
 and encode two bits classical information as
\begin{equation}\label{2m12}
\sigma^A_{00}\rightarrow 00,~ \sigma^A_{01}\rightarrow 01,
~\sigma^A_{10}\rightarrow 10, ~\sigma^A_{11}\rightarrow 11.
\end{equation}
Meanwhile, Diana and Bob, and Diana and Charlie agree on  that Bob
and Charlie can only apply unitary operations
\begin{equation}\label{2operation2}
 \sigma^X_0=I=|0\rangle\langle 0|+|1\rangle\langle1|, ~ \sigma^X_1=\sigma_x=|0\rangle\langle1|+|1\rangle\langle0|
\end{equation}
 to encode  one bit classical information as following
\begin{equation}\label{2m11}
 \sigma^X_0\rightarrow 0,~ \sigma^X_1\rightarrow 1,
\end{equation}
  respectively. Here X can be either B or C.

Suppose that Alice, Bob,  Charlie, and Diana initially share GHZ
quadruplets $|GHZ\rangle_{1234}$ and
  $|GHZ\rangle_{5678}$, where particles 1 and 5, 2 and 6, 3 and 7,
  and 4 and 8 belong to Alice, Bob, Charlie and Diana respectively.
  After senders  encode their secret messages by
applying predetermined unitary operators on $|GHZ\rangle_{1234}$,
 $|GHZ\rangle_{1234}$ becomes
$\sigma^A_{ij}\otimes\sigma^B_l\otimes\sigma^C_m|GHZ\rangle_{1234}$,
where $i,j,l,m\in \{0,1\}$.

It is easy to prove that
\begin{eqnarray}\label{2original1}
 &&\sigma^A_{ij}\otimes\sigma^B_l\otimes\sigma^C_m|GHZ\rangle_{1234}\otimes|GHZ\rangle_{5678}\nonumber\\
&=&\frac{1}{4}[\sigma^A_{ij}\Phi^+_{15}\otimes\sigma^B_l\Phi^+_{26}\otimes\sigma^C_m\Phi^+_{37}\otimes\Phi^+_{48}\nonumber\\
&&
+\sigma^A_{ij}\Phi^+_{15}\otimes\sigma^B_l\Phi^+_{26}\otimes\sigma^C_m\Phi^-_{37}\otimes\Phi^-_{48}\nonumber\\
&&
+\sigma^A_{ij}\Phi^+_{15}\otimes\sigma^B_l\Phi^-_{26}\otimes\sigma^C_m\Phi^+_{37}\otimes\Phi^-_{48}\nonumber\\
&&
+\sigma^A_{ij}\Phi^+_{15}\otimes\sigma^B_l\Phi^-_{26}\otimes\sigma^C_m\Phi^-_{37}\otimes\Phi^+_{48}\nonumber\\
&&
+\sigma^A_{ij}\Phi^-_{15}\otimes\sigma^B_l\Phi^+_{26}\otimes\sigma^C_m\Phi^+_{37}\otimes\Phi^-_{48}\nonumber\\
&&
+\sigma^A_{ij}\Phi^-_{15}\otimes\sigma^B_l\Phi^+_{26}\otimes\sigma^C_m\Phi^-_{37}\otimes\Phi^+_{48}\nonumber\\
&&
+\sigma^A_{ij}\Phi^-_{15}\otimes\sigma^B_l\Phi^-_{26}\otimes\sigma^C_m\Phi^+_{37}\otimes\Phi^+_{48}\nonumber\\
&&
+\sigma^A_{ij}\Phi^-_{15}\otimes\sigma^B_l\Phi^-_{26}\otimes\sigma^C_m\Phi^-_{37}\otimes\Phi^-_{48}\nonumber\\
&&+\sigma^A_{ij}\Psi^+_{15}\otimes\sigma^B_l\Psi^+_{26}\otimes\sigma^C_m\Psi^+_{37}\otimes\Psi^+_{48}\nonumber
\\&&
+\sigma^A_{ij}\Psi^+_{15}\otimes\sigma^B_l\Psi^+_{26}\otimes\sigma^C_m\Psi^-_{37}\otimes\Psi^-_{48}\nonumber\\
&&
+\sigma^A_{ij}\Psi^+_{15}\otimes\sigma^B_l\Psi^-_{26}\otimes\sigma^C_m\Psi^+_{37}\otimes\Psi^-_{48}\nonumber\\
&&
+\sigma^A_{ij}\Psi^+_{15}\otimes\sigma^B_l\Psi^-_{26}\otimes\sigma^C_m\Psi^-_{37}\otimes\Psi^+_{48}\nonumber\\
&&
+\sigma^A_{ij}\Psi^-_{15}\otimes\sigma^B_l\Psi^+_{26}\otimes\sigma^C_m\Psi^+_{37}\otimes\Psi^-_{48}\nonumber\\
&&
+\sigma^A_{ij}\Psi^-_{15}\otimes\sigma^B_l\Psi^+_{26}\otimes\sigma^C_m\Psi^-_{37}\otimes\Psi^+_{48}\nonumber\\
&&+\sigma^A_{ij}\Psi^-_{15}\otimes\sigma^B_l\Psi^-_{26}\otimes\sigma^C_m\Psi^+_{37}\otimes\Psi^+_{48}\nonumber\\
&&
+\sigma^A_{ij}\Psi^-_{15}\otimes\sigma^B_l\Psi^-_{26}\otimes\sigma^C_m\Psi^-_{37}\otimes\Psi^-_{48}].\nonumber\\
\end{eqnarray}
Here
\begin{equation}\label{2EPR}
    \Phi^{\pm}\equiv\frac{1}{\sqrt{2}}(|00\rangle\pm |11\rangle),
    \Psi^{\pm}\equiv\frac{1}{\sqrt{2}}(|01\rangle\pm|10\rangle)
\end{equation}
are  four Bell states (EPR pairs). By  virtue of Eq.(6), it is not
difficult to observe  that only four unitary operators
$\sigma^A_{ij}\otimes\sigma^B_l\otimes\sigma^C_m$  can correspond to
a fixed  outcome of Alice, Bob and Charlie's Bell state
measurements. For example, if  Alice obtains measurement result
$\Psi^+_{15}$, Bob $\Phi^+_{26}$, and Charlie $\Psi^+_{37}$, we can
see that only $\sigma^A_{00}\otimes\sigma^B_1\otimes\sigma^C_0$,
$\sigma^A_{01}\otimes\sigma^B_0\otimes\sigma^C_1$,
$\sigma^A_{10}\otimes\sigma^B_0\otimes\sigma^C_1$,
$\sigma^A_{11}\otimes\sigma^B_1\otimes\sigma^C_0$ can make
$|GHZ\rangle_{1234}\otimes|GHZ\rangle_{5678}$ to cause this
measurement result $\Psi^+_{15}$,  $\Phi^+_{26}$, and $\Psi^+_{37}$.
So  the capacity of the scheme in transmitting the secret messages
is 2 bits when the encoding scheme is announced publicly. For the
case $M> 3$ one can obtain the same result. That is, when the
encoding scheme is announced publicly,  the capacity of the scheme
is 2 bits, no matter how big $M$ is.

Now let us consider this problem from another point of view.
Obviously only based on Diana's Bell state measurement result, can
Diana infer the senders' secret message, as the senders outcomes
of the Bell state are announced publicly. However as a matter of
fact, Diana can only obtain four different results in a Bell state
measurement. Therefore the capacity of the scheme in transmitting
the secret messages can not beyond 2 bits.

Fortunately, if the encoding scheme (for example Eqs.(2)-(5)) is
kept in secret, the situation would be changed, as there are many
encoding schemes.  That is, if the information on the encoding
scheme is not available to the eavesdropper, there is no way for the
eavesdropper to find correct classical secret bits with a
probability more than $\frac {1}{2^{M+1}}$. Therefore in this case
the capacity is also $M+1$ bits. In other words, if the encoding
scheme is secret then other $M$ parties can transmit $M+1$ bit
classical messages to the center party.

In summary,  we have analyzed the  capacity of  a simultaneous
quantum secure direct communication  scheme between the central
party and other $M$  parties via $M+1$-particle GHZ states and
swapping quantum entanglement. It is shown that the encoding scheme
should be secret if other $M+1$ parties wants to transmit $M+1$ bit
classical messages to the center party secretly. However when the
encoding scheme is announced publicly, we prove that the capacity of
the scheme in transmitting the secret message is 2 bits, no matter
how big $M$ is.

\acknowledgments The authors sincerely thank Prof. Long Gui-Lu and
Dr. Deng Fu-Guo for  many insightful discussions and helpful
comments on the manuscript. This work was supported by Hebei Natural
Science Foundation of China under Grant Nos: A2004000141 and
A2005000140, and Key Natural Science Foundation of Hebei Normal
University.


\begin{thebibliography}{999}\footnotesize


\bibitem{BB84}  Bennett C H and  Brassard G  {\it Proc. IEEE Int. Conf. on Computers,
Systems and Signal Processing}, Bangalore, India, (IEEE, New York,
1984),  pp. 175-179
\bibitem{WZ}  Wootters W K  and  Zurek W H 1982  \emph{Nature} {\bf 299}  802
\bibitem{NC}  Nielsen M A and Chuang  I L  2000
\emph{Quantum Computation and Quantum Information}
           (Cambridge: Cambridge University Press)
\bibitem{ShorPreskill}  Shor P W and Preskill J  2000 {\it Phys. Rev. Lett. } {\bf 85}  441

\bibitem{Ekert91}  Ekert A K 1991 {\it Phys. Rev. Lett. } {\bf 67}  661
\bibitem{B92} Bennett  C H 1992 {\it Phys. Rev. Lett. } {\bf 68}  3121
\bibitem{BBM92}  Bennett C H,  Brassard G and  Mermin N D 1992 {\it Phys. Rev. Lett.} {\bf 68}  557
\bibitem{LongLiu}  Long G  L and  Liu X S 2002 {\it Phys. Rev. } A {\bf 65}  032302
\bibitem{XueLiGuo}  Xue P,  Li C F and  Guo G C 2002 {\it Phys. Rev. } A {\bf 65}  022317
\bibitem{DengLong2003}  Deng F G and  Long G L 2003 {\it Phys. Rev. }  A {\bf 68}  042315
\bibitem{DengLong2004}  Deng F G and  Long G L 2004 {\it Phys. Rev. }  A {\bf 70}  012311
\bibitem{GRTZ}  Gisin N, Ribordy G, Tittel W and Zbinden H 2002 {\it Rev. Mod. Phys.  }  {\bf 74}
145
\bibitem{SIpra60}  Shimizu K and  Imoto N 1999 {\it Phys. Rev.
}  A {\bf 60}  157
\bibitem{SIpra62}  Shimizu K and  Imoto N 2000 {\it Phys. Rev. } A {\bf 62} 054303
\bibitem{Beige}  Beige A {\it et al} 2002   {\it Acta Phys. Pol. }  A {\bf 101}   357
\bibitem{BF}  Bostr\"{o}m K and  Felbinger T 2002  {\it Phys. Rev. Lett. }  {\bf 89}  187902
\bibitem{Wojcik}  W$\acute{\rm o}$jcik A 2003  {\it Phys. Rev. Lett.}  {\bf 90}   157901
\bibitem{CaiLi} Cai Q Y and Li B W 2004 \emph{Chin. Phys. Lett.} \textbf{21} 601
\bibitem{DLL}  Deng F G,  Long G L and  Liu X S 2003  {\it Phys. Rev.}   A {\bf 68}  042317
\bibitem{DLpra69}  Deng F G and  Long G L 2004 {\it Phys. Rev.}  A {\bf 69}  052319
\bibitem{WDLLLpra71} Wang C, Deng F G, Li Y S, Liu X S and  Long G L 2005 {\it Phys. Rev. }  A {\bf 71}  044305
\bibitem{ManZhangLi} Man Z X, Zhang Z J and Li Y 2005 \emph{Chin. Phys. Lett.}
 \textbf{{22}} 18
\bibitem{ZhuXiaFanZhang} Zhu A D, Xia Y, Fan Q B and Zhang S 2006 \emph{Phys. Rev.} A \textbf{73} 022388
\bibitem{CaoSong} Cao H J and Song H S 2006 \emph{Chin. Phys. Lett.}
\textbf{23} 290
\bibitem{YZ}  Yan F L and  Zhang X Q 2004 {\it Eur. Phys. J. } B {\bf 41}  75
\bibitem{Gaozna}  Gao T 2004 {\it Z. Naturforsch.}  {\bf 59a}  597
\bibitem{GYWcp}  Gao T,  Yan F L and  Wang Z X 2005  \emph{Chin. Phys.} {\bf 14} 893
\bibitem{GYWijmpc} Gao T, Yan F L and  Wang Z X 2005   \emph {Int.  J. Mod. Phys.}
C \textbf{16} 1293
\bibitem{GYWnc}  Gao T,  Yan F L and  Wang Z X 2004 {\it Il Nuovo Cimento} {\bf 119B}  313
\bibitem{GYWjpa2} Gao T,  Yan F L and  Wang Z X 2005 \emph{J. Phys. } A \textbf{38} 5761
\bibitem{GYWcpl} Gao T,  Yan F L and  Wang Z X 2005 \emph{Chin. Phys. Lett.}  \textbf{22}
2473
\bibitem{DengLiLiZhouLiangZhou} Deng F G, Li X H, Li C Y, Zhou P,
Liang Y J and Zhou H Y,  to be published.
\end{thebibliography}
\end{document}